\documentclass[preprintnumbers,preprint,showpacs, nofootinbib]{revtex4}

\usepackage{bm}
\usepackage{latexsym}
\usepackage{amsfonts}
\usepackage{amssymb}
\usepackage{amsmath}

\newcommand{\bal}{\begin{equation}\begin{aligned}}
\newcommand{\eal}{\end{aligned}\end{equation}}

\newcommand{\ri}{r_\infty}

\begin{document}
\vspace*{4cm}

\bigskip
\title{ 
Mass and Free energy in Thermodynamics of Squashed Kaluza-Klein Black Holes
}

\author{\Large
	Yasunari Kurita\footnote{E-mail:kurita@sci.osaka-cu.ac.jp}}
   \address{  
  \bigskip
	Osaka City University Advanced Mathematical Institute,
  	Osaka 558-8585, Japan}          
  \author{\Large
        Hideki Ishihara\footnote{E-mail:ishihara@sci.osaka-cu.ac.jp} 
        }
  \address{
\bigskip 
        Department of Mathematics and Physics, Osaka City University,
  	Osaka 558-8585, Japan
  }


\begin{abstract} 
The Abbott-Deser mass, the Hamiltonian and the Komar mass of 
the 5-dimensional Kaluza-Klein black hole with squashed horizons take different values. 
Introducing a new couple of thermodynamic variables for the Komar mass, we show that 
each mass can be interpreted as a thermodynamic potential with its own natural variables, 
i.e. all masses are related to each other by the Legendre transformations. 
It is found that the new variables and the gravitational tension represent the squashing of the outer horizon.
\end{abstract}


\preprint{OCU-PHYS 265}
\preprint{AP-GR 41}

\maketitle

\section{Introduction}

In recent years, the studies on black holes in higher dimensions have attracted much attention. 
Some of these studies show that they have much more complicated and richer structure than 4-dimensional ones.
Especially, the study of Kaluza-Klein (KK) black holes is important 
in association with our apparent 4-dimensional spacetime. 
The Gregory-Laflamme instability~\cite{Gregory:1993vy} 
(see also \cite{Harmark:2007md} and references therein) and black hole/black string phase transition 
(see {\it e.g.} the reviews \cite{Kol:2004ww}) 
motivate many studies on thermodynamic aspects of KK black holes. 
Now, the thermodynamic properties and the first law for asymptotically flat KK black holes 
are widely investigated \cite{Kol:2003if}-\cite{Kastor:2007wr}.

In 5-dimensional Einstein-Maxwell theory, there is an analytic solution representing an 
electrically charged black hole with squashed horizons 
\cite{Ishihara:2005dp} as a generalization of the solution given in \cite{Dobiasch:1981vh} and \cite{Gibbons:1985ac}. 
Intriguingly, the spacetime far from the black hole is 
locally a product of the 4-dimensional Minkowski spacetime and $S^1$. 
In this sense, the black hole resides in KK spacetime and 
is worth to be named a KK black hole.

The black hole has an interesting property that various definitions of mass take different values, 
which means that the black hole gives an opportunity to investigate 
the differences in various definitions of mass. 
In this paper, we show some expressions for the first law of black hole thermodynamics 
satisfied by those masses and discuss the differences from the viewpoint of thermodynamics.

\section{Kaluza-Klein black hole}

Let us review the KK black holes  with squashed horizons~\cite{Ishihara:2005dp}, 
which is a solution of the 5-dimensional Einstein-Maxwell theory. 
The action is given by 
 \begin{eqnarray}
 I = \frac{1}{16\pi G} \int_M d^5 x \sqrt{-g} 
         \left[ R - F^{\mu\nu} F_{\mu\nu} \right]+\frac{1}{8\pi G} \int_{\partial M} 
 K\sqrt{-h}d^4x, 
 \end{eqnarray}
 where $G$ is the 5-dimensional Newton constant, $g_{\mu\nu}$ is the metric, $R$ is scalar curvature, 
 $F_{\mu\nu} = \partial_\mu A_\nu - \partial_\nu A_\mu $ is 
 the field strength of the 5-dimensional $U(1)$ gauge field $A_\mu$. 
The second term is so-called Gibbons-Hawking term, in which $h$ is determinant of the induced metric and 
 $K$ is trace of the extrinsic curvature of the boundary $\partial M$, respectively. 
The metric of the black hole is given by 
\begin{eqnarray}
 ds^2 &=& - V(\rho) d\tau^2 + \frac{B (\rho)}{V(\rho)} d\rho^2 
 + \rho^2 B (\rho) d\Omega^2  
 + \frac{r_{\infty}^2}{4 B (\rho)} \left(d\psi + \cos \theta d\phi \right)^2 ,
\label{eq:metric-Trho}
\end{eqnarray}
where $d\Omega^2 = d\theta^2 + \sin^2 \theta d\phi^2$ is the metric of the unit two-dimensional sphere and 
\begin{eqnarray}
V(\rho)=  \left( 1- \frac{\rho_{+}}{\rho}\right)\left( 1-\frac{\rho_{-}}{\rho} \right) , \ \ 
B(\rho) =   1+ \frac{\rho_0}{\rho}, \ \ 
r_{\infty} = 2\sqrt{(\rho_+ +\rho_0 )(\rho_- + \rho_0 )}. 
\end{eqnarray}
Here, the coordinate ranges are 
$0 \leq \theta <2\pi ,\   0 \leq \phi < 2\pi ,\  0 \leq \psi < 4\pi $. 
The gauge potential is given by
\begin{eqnarray}
A = \mp   \frac{\sqrt{3}}{2}  \frac{\sqrt{{\rho_+}{\rho_-}}}{ \rho_0}
  \left(1+\frac{\rho_0}{\rho} \right) d\tau.
 \end{eqnarray}

It is easy to see the apparent singularity at $\rho_{+}$ corresponds to the outer horizon of the black hole.
The inner horizon $\rho_{-}$ is analogous to that of the Reissner-Nordstr\"om (RN) black holes.
The spatial infinity corresponds to a limit $\rho \to \infty$. 
It should be noted that the shape of the horizon is a squashed sphere 
as was discussed in~\cite{Ishihara:2005dp}.
From the metric (\ref{eq:metric-Trho}), 
it is seen that the $S^1$ circle parametrized by a coordinate $\psi$ has finite size even at the spatial infinity. 
The non-trivial twisting of the $S^1$ circle fibrated over the $S^2$ base space 
leads a 4-dimensional U(1) gauge field by KK reduction.
Actually, in no horizon limit $\rho_+, \rho_- \to 0$, the black hole spacetime becomes the KK monopole 
spacetime \cite{Gross-Perry}\cite{Sorkin}. 
In the limit $\ri \to \infty$, the KK monopole becomes 5-dimensional Minkowski spacetime and 
the black hole reduces to the 5-dimensional RN black hole.
We term this limit the spherically symmetric limit.

Given the metric (\ref{eq:metric-Trho}), we can calculate various physical quantities.
The surface gravity is calculated as
\begin{eqnarray}
   \kappa_{+} = \frac{\rho_{+} -\rho_{-}}{2\rho_{+} \sqrt{{\rho_{+}}({\rho_{+} + \rho_0})}},
   \label{eq:surfacegravity}
\end{eqnarray}
which gives the Hawking temperature of the black hole $T= \kappa_+/2\pi$ \cite{Hawking:1974sw}.
We assume that the entropy of the black hole is given by the Bekestein-Hawking formula 
\begin{eqnarray}
S = \frac{A_+}{4G} = \frac{4\pi^2}{  G}  \rho_+  (\rho_+ + \rho_0) \sqrt{\rho_+(\rho_- + \rho_0 ) },
 \label{eq:entropy}
\end{eqnarray}
which is consistent with the Wald's entropy formula \cite{Wald:1993nt} \cite{Iyer:1994ys}. 
The electric charge and electrostatic potential of the black hole are also calculated as \cite{Ishihara:2005dp}
\begin{eqnarray}
  Q = 
    \pm \frac{\sqrt{3} \pi}{G} r_\infty \sqrt{\rho_{+} \rho_{-} }, \quad 
\Phi 
= \pm \frac{\sqrt{3}}{2} \sqrt{\frac{\rho_-}{\rho_+} }. 
\end{eqnarray}

\section{Mass and free energy}

There are several definitions of mass for the black hole spacetime.
Cai {\it et al.} \cite{Cai:2006td} discussed mass of the black hole defined by the counter-term method for
asymptotically locally flat spacetime \cite{Mann:1999pc}-\cite{Kraus:1999di}. 
Using the counter-term mass $M_{ct}$, they investigated the first law of black hole thermodynamics and 
suggested the existence of a new work term in the first law. 
The direct calculation reveals 
\begin{eqnarray}
 M_{ct} = M_{AD} = \frac{\pi}{2G}r_{\infty} \left( 2\rho_+ + 2\rho_- +\rho_0 \right), 
\label{eq:ct-mass-BH}
\end{eqnarray}
where $M_{AD}$ is the Abbott-Deser (AD) definition of mass~\cite{Abbott:1981ff} 
evaluated on a product $S^1$ bundle over 4-dimensional Minkowski spacetime, which is a completely 
flat spacetime\footnote{ Cai {\it et al.} showed that $M_{ct}$ equals the AD mass evaluated on a "twisted" 
$S^1$ bundle over 4-dimensional Minkowski spacetime.  
However, this background spacetime is not a solution of the vacuum Einstein equation 
and we can not obtain finite free energy and Hamiltonian of the black hole on it. 
Thus, we consider the flat background.
}.
Hereafter, we term this spacetime flat background, shortly.
Therefore, $M_{AD}$ satisfies the same equations as $M_{ct}$. 
The Komar mass is also meaningful as a mass of black holes which possess a timelike Killing vector.
Using the timelike Killing vector $\partial /\partial \tau $ normalized at the spatial infinity, 
we can calculate the Komar mass for the black hole (\ref{eq:metric-Trho}) as 
\begin{eqnarray}
   M_K  =  \frac{3\pi }{4 G} r_{\infty} \left( \rho_{+} + \rho_{-} \right) ,
   \label{eq:Komar}
\end{eqnarray}
where the integral is taken over the 3-dimensional sphere at the spatial infinity.
Smarr-type formula was shown generally in~\cite{Gauntlett:1998fz} for the Komar mass 
\begin{eqnarray}
 M_K -Q\Phi = \frac{3}{2} TS,
\end{eqnarray}
which is sometimes called integrated expression for the first law. 
This implies that we would have a differential expression for the first law using the Komar mass.
Clearly, since $M_K$ does not equal $M_{ct}$, then 
the expressions for the first law satisfied by these masses should be different.

In order to deduce the work term suggested by Cai {\it et al.}, 
we note the fact that, far from the black hole, the geometry locally looks like the black string. 
It is known that, in the case of the black p-branes or black string without twisting, 
the so-called gravitational tension and the size of the extra-dimension 
contribute to the first law~\cite{Kol:2003if}\cite{Townsend:2001rg}\cite{Harmark:2003eg}\cite{Traschen:2001pb}. 
One may think that, also in the case of the squashed KK black hole, 
these quantities contribute to the first law as a work term. 
The gravitational tension which can be applied to non-asymptotically flat spacetime was 
defined by using the Hamiltonian formalism to a foliation of the spacetime 
along asymptotically translationally-invariant spatial direction~\cite{Harmark:2004ch}. 
The definition requires some reference spacetime in order to give finite gravitational tension. 
As a reference background, we choose the flat background.
Then, using the definition given in \cite{Harmark:2004ch}, we can calculate the gravitational tension 
associated with the direction $\partial_{\psi}$ as
\begin{eqnarray}
\mathcal{T} = \frac{1}{4 G}\left(\rho_+ +\rho_- +2\rho_0\right). 
\end{eqnarray}
The conjugate variable to $\mathcal{T}$ is 
the size of the extra-dimension at the spatial infinity, $L:=2\pi r_{\infty}$. 
With these quantities, $M_{AD}$ is related to the entropy and the electric charge 
via the following expression for the first law: 
\begin{eqnarray}
dM_{AD} &=& TdS + \Phi dQ +\mathcal{T} dL.
\label{eq:first-law-AD}
\end{eqnarray}
Thus, the last term $\mathcal{T}dL$ is  
thought of as a work term suggested by Cai {\it et al.}~\cite{Cai:2006td}. 
The expression for the first law (\ref{eq:first-law-AD}) shows that 
$M_{AD}$ is a thermodynamic potential as a function of $(S,Q,L)$. 
It means that $M_{AD}$ is relevant to the thermodynamic system with natural variables $(S,Q,L)$. 
The same holds true for $M_{ct}$, because of $M_{ct}=M_{AD}$.

The free energy of the black hole is obtained from the classical Euclidean action $I_E$ as 
\begin{eqnarray}
F =T {I_E} = \frac{\pi}{2G}r_{\infty} (\rho_+ + \rho_0 ),
\end{eqnarray}
which was calculated by the traditional background subtraction method 
with the flat background~\cite{Gibbons:1976ue}.
The free energy $F$ equals one evaluated by the counter-term method.
Thus, in this case, 
the counter-term method is equivalent to background subtraction method with the flat background.
In the calculation of the free energy by the background subtraction method, we fixed the temperature, 
the electro-static potential and the size of the extra-dimension at the boundary of the spacetime. 
It follows that the free energy satisfies the following differential relation: 
\begin{eqnarray}
dF = -SdT -Qd\Phi +\mathcal{T} dL.
\label{eq:first-law-F}
\end{eqnarray}
Thus, the free energy $F$ has natural variables $(T, \Phi,L)$. 
The equations (\ref{eq:first-law-AD}) and (\ref{eq:first-law-F}) suggest the following relation with the AD mass:
\begin{eqnarray} 
F=M_{AD} - Q\Phi -TS,
\label{eq:Legendre-F-AD}
\end{eqnarray}
which can be easily shown by direct calculation. 
This relation (\ref{eq:Legendre-F-AD}) is nothing but Legendre transformation between 
$F$ and $M_{AD}$.

For asymptotically flat electrically charged black holes, it is known that Hamiltonian 
does not equal ADM mass and these two quantities differ by $Q\Phi$~\cite{Hawking:1995ap}.
The black hole we consider here is not asymptotically flat but asymptotically locally flat.
In spite of the difference in asymptotic structure, 
if we regard the AD mass as a counterpart of ADM mass, 
the same is true for the case of the squashed black hole; 
The Hamiltonian of the black hole evaluated on the flat background can be related to the AD mass as
~\cite{Hawking:1998jf}
\begin{eqnarray}
H = \frac{\pi}{2G} r_{\infty} \left( 2\rho_+ - \rho_- +\rho_0 \right)
=M_{AD} -Q\Phi.
\label{eq:Hamiltonian}
\end{eqnarray}
Often, Hamiltonian for solution without electric charge is called Hawking-Horowitz (HH) mass \cite{Hawking:1995fd}.
In the case of $Q=0$, the HH mass of the black hole is equal to the AD mass 
as is in the case of asymptotically flat black hole. 
From the relation (\ref{eq:Hamiltonian}), the first law can take the form with the Hamiltonian as 
\begin{eqnarray}
dH &=& TdS - Qd\Phi+\mathcal{T} dL.
\label{eq:first-law-flat-H}       
\end{eqnarray}
Equivalently, 
the relation (\ref{eq:Hamiltonian}) is a Legendre transformation between the Hamiltonian and the AD mass. 
The equation (\ref{eq:first-law-flat-H}) shows 
that the Hamiltonian is the thermodynamic potential with natural variables $(S, \Phi, L)$. 
The Hamiltonian is also the Legendre transform of the free energy with respect to $TS$; 
$F = H-TS$.
Therefore, 
the AD mass and the Hamiltonian are related to the free energy $F$ via Legendre transformations and 
can be interpreted as different thermodynamic potentials.

However, it can be shown that the Komar mass does not have $\mathcal{T}$ or $L$ as a natural variable. 
Now, let us obtain the differential expression for the first law by use of the Komar mass. 
In order to do so, we require that the Komar mass should be related to the free energy via Legendre transformations.
In stead of $(L, \mathcal{T})$, we introduce a couple of new variables $(\epsilon, \Sigma)$ satisfying  
\begin{eqnarray}
 F=M_K - TS -Q\Phi + \epsilon \Sigma,
\quad 
dM_K = TdS + \Phi dQ -\epsilon d \Sigma.
\label{eq:first-law-Komar}
\end{eqnarray}
Then, $(\epsilon, \Sigma)$ is determined up to a constant $C$ as
\begin{eqnarray}
\epsilon = C (2\pi r_{\infty})^2=CL^2, \quad 
\Sigma =\frac{1}{16 \pi GC r_{\infty}}  \left(\rho_+ +\rho_-+2\rho_{0}\right)=  \frac{\mathcal{T}}{2CL}. 
\end{eqnarray}
Thus, the Komar mass has this quantity $\Sigma$ as a natural variable as well as $S$ and $Q$.
The pair of variables $(\epsilon, \Sigma)$ contributes not only to the differential relation 
for the Komar mass 
but also to the following:
\begin{eqnarray}
dF = -SdT -Qd\Phi +\Sigma d\epsilon,\ 
dH = TdS -Qd\Phi +\Sigma d\epsilon, \ dM_{AD} = TdS +\Phi dQ +\Sigma d\epsilon.
\label{eq:first-law-H-AD-epsilon}
\end{eqnarray}
Therefore, the expression for the first law including the free energy, 
 the Hamiltonian or the AD mass is not unique. 
These expressions are consistent with the interpretation that $H$ or $M_{AD}$ is the thermodynamic potential 
with natural variables $(S,\Phi, L)$ or $(S, Q,L)$, because 
system with fixed $L$ 
is equivalent to that with fixed $\epsilon$ due to the relation $\epsilon \propto L^2$.

In order to clarify the relation with the case of the 5-dimensional RN black hole,
let us consider the spherically symmetric (SS) limit $r_{\infty}\to \infty$.
However, the free energy, the Hamiltonian and the AD mass evaluated on the flat background diverge in the limit.
As an alternative background, we consider the KK monopole spacetime. 
The free energy and the Hamiltonian on the KK monopole background are calculated as 
\begin{eqnarray}
\tilde{F} = \frac{\pi}{2G} r_{\infty} 
 \left(\rho_+ + \rho_0 -  \frac{r_{\infty}}{2} \right), \quad 
\tilde{H} = \frac{\pi}{2G} r_{\infty} \left(2\rho_+ -\rho_- +\rho_0 -\frac{r_{\infty}}{2} \right).
\end{eqnarray}
It is easily checked that the difference between $F$ and $\tilde{F}$ is the free energy of the KK monopole 
on the flat background; $F-\tilde{F}=\frac{\pi}{4G}\ri^2$, 
which equals the free energy calculated by the counter-term method \cite{Mann:2005cx}.
The gravitational tension calculated on the KK monopole background is 
\begin{eqnarray}
\tilde{\mathcal{T}} &=& \frac{1}{4 G} \left( \rho_+ +\rho_- + 2\rho_0 -r_{\infty}\right).
\end{eqnarray}
With this tension, 
the free energy and the Hamiltonian satisfy 
\begin{eqnarray}
d\tilde{F} = -SdT -Qd\Phi + \tilde{\mathcal{T}} dL, \quad 
d\tilde{H} = TdS -Qd\Phi + \tilde{\mathcal{T}} dL.
\end{eqnarray}
As before, $\tilde{\mathcal{T}}$ or $L$ can not be a natural variable for the Komar mass.
As in the previous case, we can obtain a couple of thermodynamic variables $(\epsilon, \tilde{\Sigma})$ satisfying
\begin{eqnarray}
\tilde{F}=M_K-TS-Q\Phi+\epsilon \tilde{\Sigma}, \quad  dM_K = TdS + \Phi dQ -\epsilon d \tilde{\Sigma}.
\label{eq:first-Komar-KKm}
\end{eqnarray}
The result is 
\begin{eqnarray}
\epsilon ={C} (2\pi\ri)^2 = {C}L^2, \quad 
\tilde{\Sigma} = 
\frac{1}{16\pi G {C} r_{\infty}}  \left(\rho_+ +\rho_-+2\rho_{0}-\ri \right)=\frac{\tilde{\mathcal{T}}}{2{C}L},
\end{eqnarray}
where $\tilde{\Sigma}$ is different from $\Sigma$ by a constant $(16 \pi GC)^{-1}$. 
Therefore, the two quantities $\Sigma$ and $\tilde{\Sigma}$ are essentially the same, 
and the differential expressions (\ref{eq:first-law-Komar}) 
and (\ref{eq:first-Komar-KKm}) are equivalent.
This is consistent with the fact that the Komar mass does not depend on the choice of background spacetime.
In the SS limit, the products $\epsilon\tilde{\Sigma}$ and $L\tilde{\mathcal{T}}$ become zero, 
and $M_K$, $\tilde{F}$ and $\tilde{H}$ 
become those of 5-dimensional RN black hole evaluated on the 5-dimensional Minkowski background.
Thus, this formulation for the squashed black hole 
includes usual thermodynamic formulation for 5-dimensional RN black hole as a limit.
In this sense, it 
is a generalized formulation of thermodynamics for 5-dimensional electrically charged static black holes.

Because $\tilde{\Sigma}$ and $\tilde{\mathcal{T}}$ vanish when the outer horizon is a perfectly-round three-sphere,
one may think that
the thermodynamic variable $\tilde{\Sigma}$ or $\tilde{\mathcal{T}}$ can be interpreted as 
a quantities representing the deformation in the shape of the horizon. 
In fact, the variables $\tilde{\Sigma}$ and $\tilde{\mathcal{T}}$ can be rewritten as 
\begin{eqnarray} 
\tilde{\Sigma}  = \frac{1}{32\pi G C}\frac{1}{R_+ \ell_+} \left({\ell_+} - R_+ \right)^2, \quad 
\tilde{\mathcal{T}} = \frac{1}{8G}\frac{ r_{\infty}}{R_+ \ell_+} \left( \ell_+ - R_+ \right)^2,
\label{eq:Sigma-squashing}
\end{eqnarray}
where we have defined new parameters $R_+$ and $\ell_+$ as follows:
 \begin{eqnarray}
R_+ := \sqrt{\rho_+(\rho_+ + \rho_0 )}, \quad {\ell_+} := \sqrt{\rho_+ (\rho_- +\rho_0)}, 
 \end{eqnarray}
which denote 
the circumference radius of the $S^2$ base space at the outer horizon and 
that of the twisted $S^1$ fibre there, respectively. 
In this way, $\tilde{\Sigma}$ and $\tilde{\mathcal{T}}$ measure the squashing of the outer horizon.

\section{Summary}

As shown in this paper, 
the Abbott-Deser mass which equals the counter-term mass, the Komar mass and the Hamiltonian 
contribute to different expressions for the first law 
and are related to each other by the Legendre transformations. 
Each mass can be interpreted as a thermodynamic potential with its own natural variables.
The consistent set of natural variables for each mass has been revealed, 
and we have obtained a more general thermodynamic formulation for electrically charged black holes
in 5-dimensional Einstein-Maxwell system.

Now, we discuss the relation between $(L, \mathcal{T})$ and 
the pair of new quantities ($\epsilon, \Sigma$).
Let us begin by considering the case of the free energy. 
In the evaluation of the free energy, the size of the extra-dimension at spatial infinity $L$ was fixed, 
so that $L$ is a natural variable of the free energy. 
In general, $L$ can be replaced by any monotonic function of $L$, say $f(L)$, as a thermodynamics variable. 
This may be trivial because thermodynamic environment characterized by fixing the size $L$ is equivalent to
that by fixing $f(L)$. 
One may write the differential relation for the free energy as 
\begin{eqnarray}
dF = -SdT -Qd\Phi + W d f(L),
\label{eq:first-law-fL}
\end{eqnarray}
where $W$ is conjugate to $f(L)$. 
Since the last work term can be rewritten as $Wdf(L) =W f'(L) dL$,  
the relation (\ref{eq:first-law-fL}) is equivalent to (\ref{eq:first-law-F}) 
if the conjugate variable $W$ satisfies 
\begin{eqnarray}
W= \frac{\mathcal{T}}{f'(L)}.
\end{eqnarray}
In this way, the work term with the form $\mathcal{T} dL$ can be replaced by $W d f(L)$.
The first law (\ref{eq:first-law-fL}) is equivalent to (\ref{eq:first-law-H-AD-epsilon}),
if $f(L) = CL^2 = \epsilon$, 
and the relation between $\Sigma$ and $\mathcal{T}$ is given as
\begin{eqnarray}
\Sigma = \frac{\mathcal{T}L}{2\epsilon}\  \propto\  \frac{\mathcal{T}}{2L}.
\label{eq:Sigma-T-relation}
\end{eqnarray}
The Hamiltonian and the AD mass are Legendre transforms of the free energy with respect to 
$TS$ or $TS+Q\Phi$ respectively, so that the expressions in (\ref{eq:first-law-H-AD-epsilon}) are 
equivalent to (\ref{eq:first-law-AD}) and (\ref{eq:first-law-flat-H}). 
Therefore, the work term in the first law for the Hamiltonian and the AD mass is not unique.

However, the quantities $\mathcal{T}$ and $\Sigma$, which are conjugate to $L$ and $\epsilon$, 
are different in the sense of thermodynamics: 
thermodynamic environment characterized by fixing $\Sigma$ is one by fixing $\mathcal{T}/2L$, as shown in 
(\ref{eq:Sigma-T-relation}). 
Thus, the pair ($\epsilon, \Sigma$) is thermodynamically different from ($L, \mathcal{T}$). 
It is natural that thermodynamic potentials suitable for different environments are different. 
The Komar mass is a thermodynamic potential for environment characterized by ($S, Q, \Sigma$), 
while the AD mass is a thermodynamic potential for that by ($S, Q, \epsilon$) or ($S, Q, L$). 
In this way, we can interpret the difference of masses from thermodynamical viewpoint.

It is interesting to investigate thermodynamic properties and stability of the black hole in each environment 
and to compare the black hole with black string. 
It will be reported in a future publication.

\begin{acknowledgments} 
Y.K. was partially supported by the Yukawa memorial foundation and is also supported by 
the 21st Century COE "Constitution of wide-angle mathematical basis focused on knots". 
This work is supported by the Grant-in-Aid for Scientific Research Fund of the Ministry of Education, Science and 
Culture of Japan No. 13135208 and No. 14540275. 
\end{acknowledgments} 


\end{document}